\documentclass[11pt]{article}

\usepackage{amsmath}
\usepackage{amssymb}
\usepackage{amsfonts}
\usepackage{latexsym}
\usepackage{graphicx}
\usepackage{latexsym}
\usepackage{color}

\usepackage{indentfirst}

\usepackage{slashed}

\usepackage[hyperindex]{hyperref}
\hypersetup{colorlinks=true,linkcolor=blue,citecolor=blue,urlcolor=blue}

\def\ln{\,\mbox{ln}\,}

\def\sTr{\,\mbox{sTr}\,}

\def\al{\alpha}
\def\be{\beta}
\def\ga{\gamma}
\def\Ga{\Gamma}
\def\de{\delta}
\def\De{\Delta}
\def\ep{\epsilon}
\def\vp{\varepsilon}

\def\ka{\kappa}
\def\la{\lambda}
\def\La{\Lambda}
\def\si{\sigma}
\def\Si{\Sigma}
\def\rh{\rho}

\def\ta{\tau}

\def\ph{\varphi}

\def\om{\omega}

\def\na{\nabla}
\def\pa{\partial}

\newcommand{\rd}{\mathrm{d}}

\DeclareMathOperator{\cx}{\square}

\def\beq{\begin{eqnarray}}
\def\eeq{\end{eqnarray}}

\newcommand{\nn}{\nonumber}

\usepackage{float} 
\usepackage{cite}  
\usepackage{titlesec}

\usepackage{ulem} 

\titleformat*{\section}{\large\bfseries}
\titleformat*{\subsection}{\normalsize\bfseries}

\begin{document}

 \begin{center}

{\Large
Renormalizable quantum field theory
\\
 in curved spacetime with external two-form 
field}

 \vskip 6mm

\textbf{Ioseph L. Buchbinder}$^{a,b,c}$
\footnote{E-mail address: \ buchbinder@theor.jinr.ru},
\ \
\textbf{Thomas M. Sangy}$^d$
\footnote{E-mail address: \   thomas.sangy@estudante.ufjf.br},
\ \
\textbf{Ilya~L.~Shapiro}$^d$
\footnote{E-mail address: \   ilyashapiro2003@ufjf.br}
%

\vskip 4mm

a) Bogoliubov Laboratory of Theoretical Physics,
\\
Joint Institute for Nuclear Research,
141980, Dubna, Russia
\vskip 1mm

b) Center of Theoretical Physics,
Tomsk State Pedagogical University,
\\
637041, Tomsk, Russia
\vskip 1mm

c) National Research Tomsk State University, 634050, Tomsk, Russia
\vskip 1mm

d) Departamento de F\'{\i}sica, ICE, Universidade
Federal de Juiz de Fora,
\\
36036-900, Juiz de Fora, Minas Gerais, Brazil

\end{center}
\vskip 6mm

\begin{abstract}

\noindent
We argue that the renormalizability of interacting quantum field
theory on the curved-space background with an additional external
antisymmetric tensor (two-form) field requires nonminimal interaction
of the antisymmetric field with quantum fermions and scalars. The
situation is qualitatively similar to the metric and torsion background.
In both cases, one can explore the renormalization group running for
the parameters of nonminimal interaction and see how this interaction
behaves in the UV limit. General considerations are confirmed by the
one-loop calculations in the well-known gauge model based on the
$SU(2)$ gauge group.
\vskip 3mm

\vskip 3mm

\end{abstract}

\vskip 2mm

\vskip 2mm

\section{Introduction}
\label{Intro}

Renormalizability is a relevant element of a successful model
of quantum field theory. Along with the possibility to perform
consistent loop calculations and control the process of removing
the high-energy (UV) divergences, renormalizability is instrumental
in the construction of phenomenologically acceptable models.This
criterion was one of the key stones in the development of the
Standard Model, which is the most successful quantum field theory.
A remarkable example of using such a criterion is the necessity to
introduce the nonminimal coupling between scalar field(s) and scalar
curvature in a semiclassical quantum gravity, when matter fields
are quantized and gravity is a classical background. This theory is
considered a relevant building block of the full quantum gravity
program; it is also widely used in early universe cosmology.

Note, however, that it is impossible to state a priori that the early
universe geometry should be an obligatory Riemann one. In particular,
we cannot exclude that the space-time of our Universe is described
not only by the metric. Other geometric fields may also exist, such
as, e.g., the torsion $T^\tau_{\,\,\al\be}$ (see e.g. a gravity model
with torsion in \cite{hehl-76} and reference therein). In this case,
the renormalizability of the semiclassical theory requires nonminimal
interaction of fermions and scalars with the completely antisymmetric
part of the torsion tensor, which is
dual to the axial vector $S^\mu = \vp^{\mu\nu\al\be}T_{\nu\al\be}$
\cite{bush85,bush90}. The nonminimal interactions of matter fields
with curvature and torsion have many physical consequences (see,
e.g., the review \cite{torsi}). From the point of view of known physics,
such as the Standard Model, the presence of background torsion
means that there can exist a weak, non-zero current
$\langle \bar{\psi}\ga^5\ga^\mu \psi\rangle$
of some fermions $\psi$ that can manifest in the laboratory,
astrophysical, or cosmological scales. This current may emerge, for
instance, owing to the nonperturbative effects in the early Universe.
Since we cannot theoretically exclude the existence of such a current,
one has to establish its coupling to matter fields using consistency
arguments and use all available experiments to draw the upper bounds
to its magnitude or, someday, detect it. There are many publications
about different aspects of this general program (see, e.g.,
\cite{AlanKost} for a review of main results).

It is clear that, for the same reasons, one can consider the current
$\big{<} 
\bar{\psi}\Si_{\mu\nu} \psi \big{>}$, 
where
\beq
\Si_{\mu\nu} = \frac{i}{2}\big( \ga_\mu\ga_\nu
-  \ga_\nu\ga_\mu\big)
\label{si}
\eeq
and the antisymmetric field $B_{\mu\nu}=-B_{\nu\mu}$,
corresponding to this current.\footnote{Note that the model of
an antisymmetric field was introduced in the works \cite{OgiPolu67}
(notoph theory) and \cite{KalbRamon}.}

Thus, it makes sense to extend the described above program of
semi-classical quantum gravity and include the field
$B_{\mu\nu}$ as part of external background. Note that the
totally antisymmetric fields or $p$-forms naturally arise
in extended supergravity models and string/brane theory (see, e.g.,
\cite{J}, \cite{O}, \cite{FVP} and references therein). In this
regard, the totally antisymmetric fields can be considered as a
part of the spacetime background.

In four dimensions, the simplest antisymmetric field is $2$-form
$B_{\mu\nu}$. It is natural that the starting point of including
this field into the program of semi-classical quantum gravity must
be a constructing consistent interaction of the $B_{\mu\nu}$ field
with the conventional fields, including scalar, fermion, and vector
ones. A form of consistent nonminimal coupling of fermions to
the $B_{\mu\nu}$-field was proposed in the work \cite{Avdeev1993}.
In the recent papers \cite{Cheshire,Cheshire2} such a coupling was
extended to curved spacetime, which allowed the study of the
renormalization of the one-loop effective action in the sector of
external fields.

In the present article, we introduce the nonminimal interaction of
the antisymmetric field $B_{\mu\nu}$ not only with fermionic
fields, but also with scalar ones. The non-minimal coupling of the
$B_{\mu\nu}$-field to fermion seems quite natural, since there is the
corresponding fermionic current (\ref{si}), but its coupling to a
scalar field may cause confusion. However, such a coupling is dictated
by renormalizability. Indeed, if there exists a fermion-scalar
Yukawa-type interaction, and there is a nonminimal coupling of
fermions to $B_{\mu\nu}$-field, then a fermionic loop on the
background of $B_{\mu\nu}$-field can lead to new divergences,
depending on the scalar and on the field $B_{\mu\nu}$.
To cancel these divergences, we should introduce
the corresponding counterterm in the scalar sector. By following
ultraviolet completion procedure, we require a multiplicative
renormalizability of the theory, and assume that the classical
Lagrangian from the very beginning must include a specific
non-minimal interactions of the scalar with the $B_{\mu\nu}$-field
with a new nonminimal coupling parameters. Then, the above divergence
can be eliminated by the renormalization of this nonminimal coupling
parameter. As we will show, these general arguments are completely
confirmed by the one-loop calculation of divergences in the simple
gauge model based on the $SU(2)$ group \cite{VT76} and containing
three nonminimal couplings of matter fields to external curvature and
$B_{\mu\nu}$-field. Since the theory under consideration is now
renormalizable, we can use the renormalization group arguments
and explore a behavior of the running couplings corresponding to
nonminimal interaction parameters.

The paper is organized as follows. In Sec.~\ref{sec2}, we give
general arguments about the renormalization structure of the
interacting field models in curved spacetime with an external
(background) antisymmetric tensor field.
Sec.~\ref{sec3} illustrates the general arguments by
directly calculating the one-loop divergences in the relatively simple
$SU(2)$ gauge model including nonminimal interactions of scalars
with external curvature and fermions with external antisymmetric
field. The calculations are done in the framework of the background field
method and proper-time technique for finding one-loop divergences.
The one-loop renormalization of the nonminimal couplings is
described.
Sec.~\ref{sec4} constructs and explores the renormalization group
equations for the nonminimal interaction parameters of
fermions and scalars with the external antisymmetric tensor field.
Dealing with the nonabelian theory and restricting the number of
fermion multiplets, we can explore the high-energy (UV) limit for
the nonminimal running couplings.
In Sec.~\ref{sec5}, we derive the trace anomaly and anomaly-induced
effective action for the massless conformal version of the theory.
Taking the low-energy (IR) limit according to the recent proposal
of \cite{AnoIntScal,AtA} and \cite{Vale2023}, we arrive at the
effective potential of the scalar and antisymmetric tensor field
in curved spacetime.
Finally, in Sect.~\ref{Conc} we draw our conclusions.

\section{Renormalizable theory with metric and two-form background}
\label{sec2}

Our starting point is the interaction of the field $B_{\mu\nu}$ with
a fermion in curved spacetime.  The free part of the fermionic action,
in this case, has the form
\beq
&&
S_f \,=\,
i \int \rd^4 x \sqrt{-g} \, \bar{\psi} \big(
\slashed{\na} - \eta B_{\mu \nu} \Si^{\mu \nu}
+ i m_f\big) \psi_{k}^{b} ,
\label{actfer}
\eeq
where $\eta$ is the new dimensionless nonminimal parameter,
$\slashed{\na} = \ga^\mu \na_\mu$, while $\nabla_{\mu}\psi$
is a spinor covariant derivative in curved spacetime. The Dirac
$\gamma$-matrices are defined in curved spacetime (see the
details of spinor analysis in curved spacetime, e.g., in \cite{OUP}).

In general, the interactions of fields with the geometric background
follow the principles of covariance, locality, and the absence of the
parameters with the inverse of the mass dimensions (see discussion,
e.g., in \cite{OUP}). In the purely metric background and scalar
field $\ph$, these requirements open the way for the nonminimal
term $\xi_1R\ph^2$, and this term has to be introduced with a
special dimensionless nonminimal parameter $\xi_1$. On the other
hand, the same principles forbid more complicated terms, such as
$R^{\mu\nu}\pa_\mu\ph\pa_\nu\ph$. Those terms that are
compatible with the above principle may emerge in
divergences and, therefore, are required for renormalizability. The
inclusion of other terms, typically, violates renormalizability.

Following the same logic, the term $\eta B_{\mu \nu} \Si^{\mu \nu}$
in Eq.~(\ref{actfer}) is allowed and we arrive at the new nonminimal
parameter $\eta$. Furthermore, as it was argued in the Introduction,
in the
scalar sector, we should introduce one more nonminimal term of the
form $\xi_2B_{\mu \nu}^2 \ph^2$ with a new nonminimal parameter
$\xi_2$. Thus, the action of a renormalizable real scalar field theory
with nonminimal coupling with external curvature and antisymmetric
should be taken in the form
\beq
S_0
&=&
\int \rd^4 x \sqrt{-g} \, \biggl\{
\frac{1}{2} g^{\mu \nu} \pa_\mu \ph\,  \pa_\nu \ph
- \frac{1}{2} m_0^2 \ph^2 + \frac12 \xi_1 R\ph^2
+ \frac12 \xi_2 \ph^2 B_{\mu \nu}^2  - \frac{1}{4!} f  \ph^4
\biggl\}\, .
\label{actscal}
\eeq

For the vector field, we require, as a necessary condition, the
preservation of gauge symmetry. In the non-Abelian case, this means
there is only minimal coupling of the vector field to gravity, such
that any nonminimal interactions are forbidden. For the Abelian gauge
field, there can be an additional possibility, but in this paper, we
will consider only a non-Abelian vector field. At this point, we
can state that the non-Abelian gauge theory with the nonminimal
terms in the scalar and fermionic sectors is expected to be
renormalizable in the matter sector.\footnote{These considerations
are evidently confirmed by the standard power counting arguments
since we have only dimensionless coupling constants. This means
that the described nonminimal interactions are consistent with the
power counting.}

It is important that the massless actions of free scalar, fermion, 
and gauge vector fields, with an additional condition $\xi_1  = 1/6$ 
for a scalar, are invariant under local conformal transformations
\beq
g'_{\mu\nu} = g_{\mu\nu}\,e^{2\si}\,,
\qquad
B'_{\mu\nu} = B_{\mu\nu}\,e^{\si}\,,
\qquad
\psi' = \psi\,e^{-\frac{3}{2}\si}\,,
\qquad
\varphi' = \varphi\,e^{-\si}\,,
\qquad
A'_\mu = A_\mu,
\label{confBg}
\eeq
where $\si = \si(x)$ and $A_\mu$ is a vector (Abelian or nonabelian)
field.  The nonminimal interactions with $B_{\mu \nu}$ in both 
(\ref{actfer}) and  (\ref{actscal}) cases, do not violate local 
conformal symmetry. The generalizations for multi-fermion or 
multi-scalar theories are straightforward. The modification, in 
these cases, is the need for special nonminimal parameters $\eta$ 
and $\xi_2$ for each of the fermion and scalar fields.

Renormalization of quantum field theory in external fields assumes
also eliminating divergences in the sector of external fields (vacuum
sector). For these purposes, we should take into consideration the
appropriate vacuum action with some parameters and renormalize
these parameters (see, e.g. \cite{OUP} and reference therein).
The same principles that we used in the analysis of the matter
sector also apply in determining the form of the vacuum action.
Using the notations introduced in \cite{Cheshire},
we arrive at
\beq
S_{\text{vac}}
&=&
S_{g} + S_B,
\label{actvac}
\eeq
where the purely gravitational part is
\beq
S_g
&=&
\int \rd^4 x \sqrt{-g} \, \Bigl\{ - \frac{1}{\ka^2} ( R + 2 \La )
+ a_1 C^2 + a_2 E_4 + a_3 \cx R \Bigr\}.
\label{actgrav}
\eeq
In this expression,
$C^2 = R_{\mu \nu \alpha \beta}^2 - 2 R_{\al \be}^2 + \tfrac{1}{3} R^2$
is the square of the Weyl tensor and
$E_4 = R_{\mu \nu \alpha \beta}^2 - 4 R_{\al \be}^2 + R^2$ is the
integrand of the Gauss-Bonnet topological invariant term.

In the vacuum $B_{\mu \nu}$-dependent part of $S_{vac}$
(\ref{actvac}), we need to consider both conformal and
nonconformal terms. The first set includes the expressions
\beq
&&
W_1 = B^{\mu \nu} B^{\al \be} C_{\al \be \mu \nu},
    \qquad
W_2 = ( B_{\mu \nu}^{2} )^{2},
    \qquad
W_3 = B_{\mu \nu} B^{\nu \alpha} B_{\alpha \beta} B^{\beta \mu},
    \nn
    \\
&&
W_4 = ( \na_\al B_{\mu \nu} )^2
- 4 ( \na_{\mu} B^{\mu \nu} )^2
+ 2 B^{\mu \nu} R_{\nu}^{\al} B_{\mu \al}
- \frac16 R B_{\mu \nu}^2.
\label{W1234}
\eeq
Each of these structures has the property
$\sqrt{-g} \, W_{i} = \sqrt{-\bar{g}} \, \bar{W}_{i}$ under the
local conformal transformation (\ref{confBg}). The nonconformal
terms are
\beq
&&
K_1 \,= \,B^{\mu\nu}B^{\al\be} R_{\mu\al} g_{\nu\be},
\qquad
K_2 \,= \, B_{\mu\nu}B^{\mu\nu}R = R B_{\mu \nu}^2,
\nn
\\
&&
K_3 \, = \,(\na_\al B_{\mu\nu}) (\na^\al B^{\mu\nu})
 \,= \,(\na_\al B_{\mu\nu})^2,
\nn
\\
&&
K_4\, = \,(\na_\mu B^{\mu\nu}) (\na^\al B_{\al\nu})
\,= \,(\na_\mu B^{\mu\nu})^2 .
\label{K123}
\eeq
It is easy to see that $W_4$ is a specific linear combination of
the terms (\ref{K123}). On top of this, we have to include the
possible total derivative terms, described in \cite{Cheshire2},
\beq
N_1\,=\,\cx \big(B_{\mu\nu}\big)^2\,,
\quad
N_2 \,=\,\na_\mu \big[B^{\mu\nu} \big(\na^\al B_{\al\nu}\big)\big]
\quad
\mbox{and}
\quad
N_3 \,=\,\na_\mu \big[B_{\al\nu}\big(\na^\al B^{\mu\nu} \big)\big]\,.
\label{N123}
\eeq
Following the general analysis of divergences in curved spacetime
given in \cite{tmf} (see also \cite{OUP} for a simplified
consideration), at the one-loop level, there may be only
conformal (\ref{W1234}) and total derivative  (\ref{N123})
divergences \footnote{It was shown in \cite{tmf} that one-loop
divergences in any classical conformal invariant theory
of scalar, spinor, and vector fields are conformal invariant at
one-loop level both in matter and vacuum sectors. Although
analysis in \cite{tmf} concerned
only theories in an external gravitational field, it can be extended
to include $B$-field background.}. This expectation
has been confirmed by direct calculations of the fermionic loop
\cite{Cheshire}. The nonconformal terms (\ref{K123}) are
generated in the finite one-loop corrections, which can be seen
by integrating trace anomaly \cite{Cheshire2}.
Starting from the second loop, the nonconformal terms are also
expected in the divergences, but this is beyond the approximation
used here. In the next section, we check the described general
statements by performing one-loop calculations in the simple model
based on the $SU(2)$ gauge group.

\section{One-loop renormalization of the curved spacetime
SU(2) gauge model with external antisymmetric field}
\label{sec3}

In this section, we derive the one-loop divergences for an $SU(2)$
gauge model, with Yang-Mills filed $A^{a}_{\mu}$, several copies
of Dirac fermion $\psi_k^a$ and a real scalar field $\ph_a$, both in
the adjoint representation of the gauge group, on the background
of spacetime metric $g_{\mu \nu}$ and an antisymmetric tensor field
$B_{\mu \nu}$.
The original flat-space version of this model at $B_{\mu \nu} =0$ was
used in 1976 in Ref.~\cite{VT76}to explore asymptotic freedom. Our
main focus will concern the issues related to the non-minimal
interaction of $B_{\mu\nu}$-field with scalar and spinor fields.

\subsection{Description of the model}

The theory under consideration is described by the action
\beq
S = S_{\text{vac}} +S_{\text{mat}},
\label{fullact}
\eeq
where $S_{vac}$ is an action of external fields (\ref{actvac}).
The action $S_{g}$ is given by (\ref{actgrav}) and $S_{B}$
has the form
\beq
S_{B}=\int \rd^4 x \sqrt{-g} \, \biggl\{ \frac12 ( \tau W_4 + \la W_1 )
- \frac12 M^2 B_{\mu \nu}^2 - \frac14 ( f_2 W_2 + f_3 W_3 ) \biggr\},
\label{actB}
\eeq
where we used the definitions (\ref{W1234}), $\tau$, $\la$, $f_2$
and $f_3$ are the nonminimal dimensionless parameters in the
vacuum sector, and $M$ is the mass of the background field
$B_{\mu\nu}$. The matter action $S_{mat}$ is written as follows
\beq
S_{\text{mat}}
&=&
\int \rd^4 x \sqrt{-g} \, \biggl\{- \frac{1}{4} G_{\mu \nu}^{a} G^{a \mu \nu}
\,+\,
i \sum_{k=1}^{s} \bar{\psi}_{k}^{a} ( \slashed{D}^{ab}
- \eta B_{\mu \nu} \Si^{\mu \nu} + i m_{f} \delta^{ab}
- h \varepsilon^{acb} \ph^c ) \psi_{k}^{b}
\nn
\\
&&
+ \, \frac{1}{2} g^{\mu \nu} ( D_{\mu} \ph )^{a} ( D_{\nu} \ph )^{a}
- \frac{1}{2} ( m_{s}^{2} - \xi_{1} R
- \xi_2 B_{\mu \nu}^{2} ) \ph^a \ph^a
\,- \,\frac{1}{4!} f ( \ph^{a} \ph^{a} )^{2} \biggr\}.
\label{actmat}
\eeq
Here $a,b=1,2,3$, the $\xi_1$, $\xi_{1,2}$ are the nonminimal
parameters, $m_s$ and $m_f$ are masses of scalar and spinor
fields, $f$ and $h$ are scalar and Yukawa coupling constants.
The Yang-Mills strength tensor is defined as
$G^{a}_{\mu\nu}= \pa_{\mu} A^{a}_{\nu}- \pa_{\nu} A^{a}_{\mu} + g\vp^{abc}A^{b}_{\mu}A_{\nu}^{c}$, where $\vp^{abc}$ and $g$
are the structure constants of the $SU(2)$ group and the gauge
coupling constant, respectively.
Gauge covariant derivatives are given in the form
\beq
(D_{\mu}\ph)^{a} = \pa_{\mu}\ph^{a} + g\vp^{abc}A^{b}_{\mu}\ph^{c}
\quad
\mbox{and}
\quad
(D_{\mu}\psi)^{a} = (\nabla_{\mu}\psi)^{a}
+ g \varepsilon^{abc}A^{b}_{\mu}\psi^{c}\,.
\eeq

The action (\ref{fullact} is invariant under general coordinate
transformations. As we already mentioned above, at zero masses
and at $\xi_{1}=1/6$, for arbitrary $\xi_2$ and $\eta$, the actions
$S_{B}$ and $S_{mat}$ are also invariant under the local
conformal transformations (\ref{confBg}). The tensor
$\Sigma^{\mu\nu}$, defined using the curved space $\ga$-matrices,
transforms as $\Sigma^{'\mu\nu}=e^{-2\si}\Sigma^{\mu\nu}$.

\subsection{One-loop divergences}

To compute the one-loop divergences, we use the background field
method (see, e.g., \cite{OUP}) and the proper-time Schwinger-DeWitt
technique \cite{DeWitt65}, \cite{BarVil}.

Following the background field method, we split the initial matter
fields into the background and quantum ones as follows
\beq
\ph^{a} \longrightarrow \ph^{a} + \si^{a},
\qquad
A_{\mu}^{a}
\longrightarrow A_{\mu}^{a} + B_{\mu}^{a},
\qquad
\psi_{k}^{a} \longrightarrow \psi_{k}^{a} + \chi^{a}_{k}.
\eeq
Here $\ph^{a}$, $A_{\mu}^{a}$ and $\psi_k^{a}$ denote classical
fields, while $\si^{a}$, $B_{\mu}^{a}$ and $\chi_{k}^{a}$ are the
quantum counterparts, respectively.

For calculating the one-loop effective action, one needs only the
quadratic in the quantum fields part $S^{(2)}$ of the matter action
together with a suitable covariant gauge-fixing term $S_{\text{GF}}$
for quantum fields. Such a quadratic action has the form
\beq
S^{(2)} + S_{\text{GF}}
\,\,=\,\,
\frac12 \int \rd^4 x \sqrt{-g}\,\,
\big( \sigma^{a} \,\,  B^{a \mu} \,\, \bar{\chi}^{a}_{k} \Big)
\, \textbf{H}\,
        \begin{pmatrix}
            \si^{b} \\
            B^{b}_{\nu} \\
            \chi_{l}^{b}
        \end{pmatrix},
\eeq
where the bilinear operator $\textbf{H}$ has the following matrix
elements (hear and later, $\ph^2 =  \ph^c \ph^c$)
\beq
H_{11} &=& -\,\de^{ab} \cx - ( m^{2}_{s} - \xi_{1} R
- \xi_{2} B_{\mu \nu}^{2} ) \de^{ab}
- \frac{f}{6} ( \ph^2 \de^{ab}
+ 2 \ph^{a} \ph^{b} ),
\nn
\\
H_{12} &=&  g \vp^{acb} \ph^{c} \na^{\nu}
+ 2 g \vp^{acb} ( \na^{\nu} \ph^{c} ),
\qquad
H_{13} = 2 i h \vp^{acb} \bar{\psi}_{l}^{c},
\nn
\\
H_{21} &=&  g \vp^{acb} \ph^{c} \na_{\mu}
- g \vp^{acb} (\na_{\mu} \ph^{c} ),
\qquad
H_{22} = \de^{ab} ( \de^{\nu}_{\mu} \cx
- R^{\nu}_{\mu} )
+ g^2 ( \ph^2 \de^{ab} - \ph^{a} \ph^{b} )
\de^{\nu}_{\mu},
\nn
\\
H_{23} &=& -\,2ig \vp^{acb} \bar{\psi}_{l}^{c} \ga_{\mu},
\qquad
H_{31} = 2 i h \vp^{acb} \psi_{k}^{c},
\qquad
H_{32} = -2ig \vp^{acb} \ga^{\nu} \psi_{k}^{c},
\nn
\\
H_{33} &=& 2 i \de_{kl} \left[ \de^{ab} ( \slashed{\na}
- \eta B_{\mu \nu} \Si^{\mu \nu} + i m_{f} ) - h \vp^{acb} \ph^{c}
\right].
\eeq

The one-loop correction to the effective action is given by
$\tfrac{i}{2} \sTr \log \mathbf{H}$. Here, the symbol $\sTr$
means a supertrace, which means a positive sign in the bosonic
and negative in the fermionic sectors. It is convenient to
introduce a conjugate operator $\mathbf{H}^{*}$,
\beq
    \textbf{H}^* =
    \begin{pmatrix}
        - \,\de^{bc}\,\, & 0 & 0 \\
        0 & \,\de^{bc} \de^{\la}_{\nu} \, & 0 \\
        0 & 0 & - \frac{i}2 \de^{bc}( \slashed{\na} - i m_f )
    \end{pmatrix}.
\label{noMA}
\eeq
Therefore, the one-loop contribution to the effective action
is written as follows
\beq
\frac{i}{2} \sTr \log \mathbf{H}
\,=\, \frac{i}{2} \sTr \log \mathbf{H} \mathbf{{H}}^*
- \frac{i}{2} \sTr \log \mathbf{H}^*.
\label{H*}
\eeq
The term $-\frac{i}{2}\sTr \log{H}^*$ does not depend of
$B_{\mu\nu}$. Its divergent part is well known and has a form
of action (\ref{actgrav}) with some concrete coefficients at the
geometrical invariants (details of calculations are given, e.g., in
\cite{book}, \cite{OUP}).   Thus, one can forget for
a moment about the second term in the r.h.s. of
(\ref{H*}) and focus on the operator
\beq
\textbf{H} \textbf{H}^{*}
\,=\,
\mathbf{1} \cx
+ 2 \mathbf{h}^\al \na_\al
+ \mathbf{\Pi}.
\label{opmin}
\eeq
The explicit forms of the operators $\mathbf{1}$, \
$\mathbf{h}^\al$ \ and \
$\mathbf{\Pi}$ can be found in the Appendix, together
with other details of the one-loop calculation.

The operator (\ref{opmin}) has a standard form to which the
Schwinger-DeWitt technique \cite{DeWitt65} (see also
\cite{BarVil} for further developments) is fully applicable,
and we can use this technique for calculating the divergences of
the corresponding effective action. The general expression for
the relevant part is
\beq
\bar{\Ga}^{(1)}_{\text{div}}
 \,=\, -\,\frac{\mu^{n-4}}{\ep} \int \rd^n x \sqrt{-g}
 \,  \, \text{str} \bigg( \frac{1}{2}\mathbf{P}^2
 + \frac{1}{12} \mathbf{S}_{\rh \si}^2 \bigg),
\label{a2}
\eeq
where
\beq
&&
\mathbf{P}
\,= \,\mathbf{\Pi} + \frac{\mathbf{1}}{6} R
- \na_\al \mathbf{h}^\al - \mathbf{h}_\al^2,
\nn
\\
&&
\mathbf{S}_{\al \be} \,= \,
[\na_\be, \na_\al ] \mathbf{1} + \na_\be \mathbf{h}_\al
- \na_\al \mathbf{h}_\be
+ \mathbf{h}_\be \mathbf{h}_\al
- \mathbf{h}_\al \mathbf{h}_\be.
\label{PS}
\eeq
As a result, the final expression for the divergences has the form
\beq
\bar{\Ga}^{(1)}_{\text{div}}
\,= \, \bar{\Ga}^{(1)}_{\text{div},\,1} ( g )
+ \bar{\Ga}^{(1)}_{\text{div}, \, 2}
+ \bar{\Ga}^{(1)}_{\text{div}, \, 3}.
\label{eq:Ga1Div1}
\eeq
Here $\bar{\Ga}^{(1)}_{\text{div},\,1}(g)$ is the the
well-known metric-dependent vacuum contribution (see,
e.g., \cite{OUP}), which we include here for completeness.
Furthermore,
\beq
&&
\bar{\Ga}^{(1)}_{\text{div}, \, 1}
\,=\,
- \,\,\, \frac{\mu^{n-4}}{n-4} \int \rd^n x \sqrt{-g}
\biggl\{ \om C^2 \,+\, b  E_{4}
\,+ \,c_\xi \cx R
\,+\, \be_\xi R^2
\nn
\\
&&
\qquad \qquad
+ \,\,\frac{1}{(4\pi)^2}\Big[
\big(s m_{f}^{2}  - 3 m_{s}^{2} \tilde{\xi}_{1} \big) R
\,+\, \frac{3}{2} m_{s}^{4}
- 6s m_{f}^{4}\Big] \biggr\},
\label{Gama1}
\eeq
where $\ep = (4 \pi)^2 (n - 4)$ and
$\tilde{\xi}_1 = \xi_1 - \tfrac{1}{6}$,
\beq
&&
\om = \frac{13+6s}{40}
\,,\qquad
b =  -  \frac{63+ 11 s}{120}
\,,\qquad
\be_\xi = \frac{3}{2} \tilde{\xi}^2_{1}
\nn
\\
&&
c_\xi \,=\, \frac{6s - 17}{60} - \frac{1}{2} \tilde{\xi}_{1}
\,=\, c - \frac{1}{2} \tilde{\xi}_{1}.
\label{ombc}
\eeq

The second term includes all ``pure'' $B_{\mu\nu}$-dependent
(vacuum) terms \cite{Cheshire,Cheshire2}
\beq
\bar{\Ga}^{(1)}_{\text{div}, \, 2}
&=&
-\,\frac{\mu^{n-4}}{\ep} \int \rd^n x \sqrt{-g} \,
\biggl\{ 4s \eta^2 W_1
- \biggl( 8s \eta^4 - \frac{3}{2} \xi_2^2 \biggr) W_2
+ 32s \eta^4 W_3 - 4s \eta^2 W_4
\nn
\\
&&
+\,3 \xi_2 \tilde{\xi}_1 K_2 -\frac12 \big( 6 m_s^2 \xi_2
- 48 s \eta^2 m_f^2 \big) B_{\mu \nu}^2
- \frac{\xi_2}{2} N_1
- 8s \eta^2 \left( N_2 - N_3 \right) \biggr\}.
\label{GaBvac}
\eeq
It is easy to see that this expression is a sum of the
fermionic contribution \cite{Cheshire} with multiplicity
$3s$ and an extra contribution of scalar fields.

Finally, there are the divergences in the matter fields sector,
\beq
\bar{\Ga}^{(1)}_{\text{div}, \, 3}
&=&
-\,\frac{\mu^{n-4}}{\ep} \int \rd^n x \sqrt{-g}
\, \biggl\{
\frac{(21 - 8 s)g^2}{6}
\,G_{\mu \nu}^{a} G^{a \mu \nu}
+ \frac{1}{2} ( 8 s h^2 - 8 g^2 ) ( \na_\mu \ph^a )^2
\nn
\\
&-&
\frac{1}{2} \biggl[ 48 s h^2 m^2_{f}
- \biggl( \frac{5}{3} f - 4 g^2 \biggr) m^2_{s} \biggr] \ph^2
+ (4\pi)^2 \be_\tau  \cx \ph^{2}
\nn
\\
&+&
 \frac12 \biggl[ \biggl( 4 g^2  -\frac53 f \biggr) \tilde{\xi}_1
- \frac43 g^2 + \frac{4}{3} s h^2 \biggr] R \ph^2
+\,\frac{1}{2}
\biggl( 32 s \eta^2 h^2 - \frac{5f}{3} \xi_{2} + 4 \xi_{2} g^2 \biggr)
B_{\mu \nu}^{2} \ph^2
\nn
\\
&-&
\frac{1}{4!} \biggl( -\frac{11}{3} f^2 + 8 g^2 f - 72 g^4 + 96 s h^4
\biggr) \big(\ph^2\big)^2
\nn
\\
&+&
i \sum_{k=1}^{s} \bar{\psi}_{k}^{a} \Bigl[ 2 \de^{ab}
( h^2 + 2 g^2 ) \slashed{\na}
+ 2 h ( h^2 - 6 g^2 ) \vp^{acb} \ph^c - 4 i ( h^2 - 4 g^2 )
\de^{ab} m_f \Bigr] \psi_k^b  \biggr\}.
\qquad
\label{eq:Ga1Div4}
\eeq
It is worth noting that the coefficient $c_\xi$ in (\ref{ombc}) and
$\be_\tau$ in the last expression (\ref{eq:Ga1Div4}) are subjects of
regularization ambiguity. We refer the interested reader to
Ref.~\cite{AnoIntScal} for the discussion of this ambiguity in
dimensional and Pauli-Villars cases.

An important detail in the one-loop result is the absence
of a divergent term of the form
$\bar{\psi}_{k}^{a} B_{\mu \nu} \Si^{\mu \nu} \psi_{l}^{b}$
(see also report on the additional verification in Appendix B).
We note that such a term is present in the classical action (as otherwise
the theory cannot be renormalizable) and is allowed by all the
symmetries of the model, so its absence cannot be attributed to
symmetry constraints. Instead, its absence follows from purely
algebraic reasons. Specifically, the immediate reason is that the
first two entries in the third row of the matrix operator
$\hat{h}^{\al}$ vanish. The analysis of the calculations leading
to this output shows that it is because of the identity of the
gamma matrices
$\ga_{\al} \Sigma^{\mu \nu} \ga^{\al} = 0$. This identity
holds independently of the representation of gamma matrices,
choice of the gauge group, or representation of the fermions
and scalars. Thus, the structure of renormalization on the
background of metric and field $B_{\mu \nu}$ is expected
to be universal and independent of the model.

As a consistency check, consider three special cases of the model.
\textit{i)} In the absence of the field $B_{\mu \nu}$, we recover the
divergences of model \cite{VT76}.
\textit{ii)} On the other hand, switching off the background fermion
and scalar fields, setting $h=f=0$ and also choosing $s=1$,
the divergent part of the one-loop effective action is three times the
result of \cite{Cheshire}, which corresponds to the dimension of the
adjoint representation of the $SU(2)$ gauge group.
\textit{iii)} Suppose the classical
theory possesses conformal symmetry described above. As it
should be \cite{tmf,OUP}, this symmetry holds in the coefficients
of the poles in the divergences, in the limit $n\to 4$. The last means
the nonconformal terms are either mass-dependent, or proportional
to $\tilde{\xi}_{1}$, or total derivatives.

The main new element is the divergence of the
$B_{\mu \nu}^{2} \ph^2$-type with the coefficient $32 s \eta^2 h^2$.
This detail shows that the introduction of the nonminimal term
$\xi_2 B_{\mu \nu}^2 \ph^2$ in the action is necessary for
the consistency of the theory. Without this term, there is no
multiplicative renormalizability. Furthermore, the introduction of
this term in the classical action results in a divergent vacuum
contribution of the type
$\xi_2 \tilde{\xi}_1 K_2 = \xi_2 \tilde{\xi}_1 R B_{\mu \nu}^2$
in Eq.~(\ref{GaBvac}). The correctness of the result is confirmed
by the fact that this term vanishes in the conformal case, when
$\xi_1 = 1/6$.

\section{Renormalization group and running couplings}
\label{sec4}

The renormalization group equations can be derived in a standard
way using the expression for divergences (\ref{eq:Ga1Div1}) (the
details of renormalization transformations are collected in
Appendix B). The equations for the conventional running couplings
constants $g,\, h,\, f$ are the same as in flat spacetime \cite{VT76},
\beq
&&
\mu \frac{dg^2}{d \mu}
\,=\, (n-4)g^2  - \frac{2}{3(4\pi)^2}(21 - 8 s) g^4 ,
\label{betag2}
\\
&&
\mu \frac{dh^2}{d \mu}
\,=\, (n-4)h^2  + \frac{1}{(4\pi)^2}
\left[8 ( 1 + s ) h^4 - 24 h^2 g^2  \right] ,
\label{betah2}
\\
&&
\mu \frac{df}{d \mu}
\,=\, (n-4)f  + \frac{1}{(4\pi)^2}
\biggl( \frac{11}{3} f^2 - 24 g^2 f + 16 s h^2 f + 72 g^4 - 96 s h^4 \biggr)\,.
\label{betaf}
\eeq
These equations have the universal form
\beq
\frac{d\la}{dt} = (n-4)\la + \be_\la,
\quad
\mbox{where}
\quad
t = \ln  \frac{\mu}{\mu_0}
\label{betadef}
\eeq
and $\la = (g^2,\,h^2,\,f)$. The reduced equations of the form
$d\la/dt = \be_\la$ will be used below to explore the UV limit
for all parameters, including the running of the nonminimal
parameters related to the new field $B_{\mu\nu}$. However,
we shall write all renormalization group equations in the complete
form similar to (\ref{betag2}), (\ref{betah2}), and (\ref{betaf}),
for the sake of completeness.
The equations for the nonminimal parameter $\xi_1$ have been
discussed in detail in many papers and the book \cite{book}, so
we skip this part and go to the ones for $\eta$ and $\xi_2$.
The corresponding equations are written as follows
\beq
&&
\frac{d \eta^2}{d t}
\,=\,\frac{1}{( 4 \pi )^2} ( 4 h^2 + 8 g^2 ) \eta^2\,,
\label{betaeta}
\\
&&
\frac{d \xi_2}{d t}
\,=\,
\frac{1}{(4\pi)^2}
\bigg[\biggl( \frac{5}{3} f
+ 8 s h^2
- 12 g^2\biggr) \xi_2
-32 s \eta^2 h^2\bigg].
\label{betaxi2}
\eeq

The last two equations have direct physical meaning.
Eq.~(\ref{betaeta}) tells us that if there is no coupling of fermions
with the external two-form (i.e., $\eta =0$), the theory is
renormalizable in the
fermionic sector. This is a consequence of the fact that the
interaction with the antisymmetric field is purely nonminimal.
On the other hand, as far as $\eta  \neq 0$, the nonzero parameter
$\xi_2$ becomes a necessary condition for renormalizability. If
this parameter is vanishing at the reference scale $\mu_0$, the
running (\ref{betaxi2}) makes it nonzero at other scales.

To explore the running of the nonminimal parameters
$\eta$ and $\xi_2$, we need the solutions for the couplings, i.e.,
\beq
g^2 (t) \,=\, \frac{g_{0}^{2}}{1 + b^2 g_{0}^{2} t},
\qquad
b^2 = \frac{1}{(4\pi)^2}\,\biggl(14 - \frac{16}{3} s\biggr).
\label{AFg}
\eeq
for Eq.~(\ref{betag2}). In the cases when  $s=1$ and $s=2$, there
is an asymptotic freedom regime in the model under consideration,
and we can study the UV asymptotic behavior of all remaining
effective charges.  For $s\geq 3$, one can explore only the
IR (low-energy) limit in the massless case, which we will not
consider here.
Following  \cite{VT76}, let us use special solutions of the
equations for Yukawa and self-scalar couplings in the form
\beq
h^2(t) = k_1 g^2(t),
\qquad
f(t) = k_2 g^2(t),
\label{k12}
\eeq
where $k_{1,2}$ are some constants. Using Eqs.~(\ref{betah2})
and ~(\ref{betaf}), one can easily get their values,
\beq
&&
k_{1} =  \frac{15 + 8s}{12 ( 1 + s )},
\nn
\\
&&
k_{2} = \pm \sqrt{\frac{97}{22}}
\quad  \text{for } s=1
\quad
\mbox{and}
\quad
k_{2} = -\frac{31}{33} \pm \frac{\sqrt{2429}}{11}
\quad
\text{for } s=2.
\label{eq:k's}
\eeq

Substituting to Eq.~(\ref{betaeta}), the solution for the effective
charge $\eta$ related to the fermionic coupling to the two-form
field
\beq
\eta^2 ( t ) \,=\, \eta_{0}^{2}
\left( 1 + b^2 g_{0}^{2} t \right)^{\frac{k_{1}+2}{4 \pi^2 b^2}}.
\eeq
This means the nonminimal interaction becomes stronger in the UV
and weaker in the IR. It is worth noting that this asymptotic behavior
is the same as the one in the case of external antisymmetric torsion
(or dual to it axial vector)
\cite{bush85,bush90}. On the other hand, the arguments concerning
the universality of the sign of the beta function \cite{torsi} remain
valid. This means that the running of the type (\ref{AFg}) should
be expected in any gauge theory with the asymptotic freedom
behaviour for all coupling constants.

Using the special solution (\ref{k12}), the Eq.~\eqref{betaxi2}
becomes
\beq
\frac{d \xi_{2}}{d t}
= \frac{1}{(4\pi)^2}\left(c_1 g^2\xi_{2} - c_2\eta^2 g^2 \right),
\eeq
with
\beq
c_1 = \frac{5}{3} k_2 + 8 s k_1 - 12,
\qquad
c_2 = 32 s k_1.
\eeq
The solution to this equation is
\beq
\xi_{2} ( t )
&=&
\biggl[
\xi_{2} ( 0 ) - \frac{c_{2} \eta_{0}^2}{c_{1}
- 4k_{1}-8} \biggr]
\left( 1 + b^2 g_0^2 t \right)^{\frac{c_1}{16 \pi^2 b^2}}
\nn
\\
&&
+ \,\,\frac{c_{2} \eta_{0}^2}{c_{1} - 4k_1 - 8}
\big(1 + b^2 g_{0}^{2} t \big)^{\frac{2+k_{1}}{4 \pi^2 b^2}}.
\eeq
Evaluating the coefficients numerically, we find
\beq
c_{1} = -0.833694
\quad  \text{for } \quad s=1
\qquad
\mbox{and}
\qquad
c_{1} = 7.67953
\quad
\text{for}
\quad s=2.
\eeq
For $s = 1$, the negative value of $A$ implies that the first term
in the solution for $\xi_2(t)$ vanishes in the UV limit. However,
in both cases, the second term dominates asymptotically, and thus
we find, in the UV,
\beq
\left| \xi_{2} ( t ) \right| \,\longrightarrow \, \infty.
\eeq
All in all, the nonminimal interaction of both fermions and scalars
becomes stronger at higher energies. From the physical side, if the
$B_{\mu\nu}$ background exists, this behavior may help to explain
why this field evade the high-precision low-energy experiments.

\section{Trace anomaly and anomaly-induced action }
\label{sec5}

As the last part of our analysis, consider the trace anomaly,
anomaly-induced action and the low-energy (IR) limit in the
theory under consideration. As usual, the conformal invariance
of the classical theory (i.e., massless and with $\xi_1= 1/6$)
breaks down due to quantum corrections, yielding the conformal
anomaly.

\subsection{Anomaly}
\label{sec51}

In the one-loop approximation, the vacuum part does not depend
in the field's interaction, and hence the situation in the theory with
scalars and gauge vector fields
does not change qualitatively compared to the pure fermionic
theory, considered in the recent previous work \cite{Cheshire2}.
Therefore, we should focus on the new part, i.e., on the
scalar-$B_{\mu\nu}$-metric sector of the anomaly. In this case,
we can use the approach of the recent papers \cite{AnoIntScal},
also \cite{AtA}, and \cite{Vale2023}. In these works, it was
shown how to perform an IR limit in the covariant nonlocal
form of the anomaly-induced effective action, which turns out the
shortest way to arrive at the effective potential of scalar and
torsion fields and of their combination \cite{Vale2023}. So, in the
present section, we will make a similar analysis as in these works,
but for the combination of $B_{\mu\nu}$ and metric background.

An important consequence of that is that we can ignore purely metric
and $B_{\mu\nu}$-dependent terms in the anomaly and purely
scalar-dependent terms in the anomaly (considered in
\cite{AnoIntScal}) is that in the expression for divergences
(\ref{eq:Ga1Div4}), there are no total derivative terms remaining.
Therefore, we may restrict consideration to the mixed
scalar-$B_{\mu\nu}$-metric part of the classical action
$S=S(g_{\mu\nu},\ph,B_{\mu\nu})$. At zero mass and
$\xi_1=1/6$, this action satisfies the conformal Noether identity
\beq
\mathcal{T}
\,=\,
\frac{1}{\sqrt{-g}}
\bigg(
\ph \,\frac{\de S}{\de \ph}
- \,2 g_{\mu\nu}\,
\frac{\de S}{\de g_{\mu\nu}}
-\,B_{\mu\nu}\,
\frac{\de S}{\de B_{\mu\nu}}
\bigg)
\,=\, 0.
\label{Noether_ident}
\eeq

The conformal anomaly is derived, using the divergences, in a
standard way \cite{duff77} (see,  e.g., \cite{OUP} for a simplified
approach and full details). Using divergences (\ref{eq:Ga1Div1})
one gets in the scalar sector together with the ``pure background''
terms,
\beq
&&
\langle \mathcal{T} \rangle
\,=\, -\,
\biggl[b  E_4 \,+ \, Y  \,+ \, \,c \cx R \,+ \, \,\be_\tau \cx \ph^2
\,-\,\frac{\xi_2}{2} N_1
- 8s \eta^2 \left( N_2 - N_3 \right)\biggr],
\label{Tbetas}
\eeq
where we used definitions (\ref{N123}) and the notation
\beq
&&
Y \, = \,\om C^2 \,+\, \be_\la W_1\,+\, \be_\tau W_4
\,+\, \be_\la W_2\,+\, \be_{f_3} W_3
\nn
\\
&&
\qquad
\,+\, \frac12\,\ga
\Big[ \big( \na_\mu \ph^a \big)^2 + \frac16\,R \ph^2\Big]
\,+\, \frac{1}{4!}\,\tilde{\be}_f\, \big(\ph^2\big)^2
\,+\,\frac{1}{2}\,\tilde{\be}_{\xi_2}
B_{\mu \nu}^{2} \ph^2.
\label{Y}
\eeq
The renormalization group functions in (\ref{Y}) are
\beq
&&
\ga = \frac{1}{(4\pi)^ 2}\,\big(8sh^2 - 8 g^2\big) ,
\nn
\\
&&
\tilde{\be}_f = \frac{1}{(4\pi)^ 2}\,
\biggl(\frac{11}{3} f^2 - 8 g^2 f + 72 g^4 - 96 s h^4\biggr),
\nn
\\
&&
\tilde{\be}_{\xi_2}
\,=\,
\frac{1}{(4\pi)^ 2}\,\biggl( 32 s \eta^2 h^2 - \frac{5}{3}\,f \xi_2
+ 4  g^2\xi_2 \biggr)\,,
\nn
\\
&&
\be_\tau = \frac{1}{18 (4\pi)^ 2}\,
\big(f + 12 g^2 - 12 s h^2\big),
\label{gambetas}
\eeq
and
\beq
&&
\be_\tau \,=\,-\,\frac{4s \eta^2}{(4\pi)^2}
\,,\qquad
\be_\la \,=\,\frac{4s \eta^2 }{(4\pi)^2}\,,
\nn
\\
&&
\be_{f_2} \,=\, -\, \frac{1}{(4\pi)^2}\,
\biggl( 8s \eta^4 - \frac{3}{2} \xi_2^2 \biggr),
\qquad
\be_{f_3}\,=\,\frac{32s \eta^4 }{(4\pi)^2}\,.
\label{betalata23}
\eeq

Let us note that some of the expressions differ from the
beta functions defined in the previous section, this is the
reason for notations with tildes.
As always, the terms in the anomaly can be divided into three groups.
The first one is formed by the real conformal invariants ($c$-terms),
collected in (\ref{Y}). Those include $C^2$, terms with $W_k$, and
the scalar conformal terms. The second group of terms includes one
topological term $E_4$ and the third is formed by total derivatives.

\subsection{Derivation of the anomaly-induced action }
\label{sec52}

The integration of the anomaly consists in establishing the
effective action $\Ga_{\text{ind}}$ which satisfies the anomalous
quantum version of the Noether identity (\ref{Noether_ident})
\beq
\frac{1}{\sqrt{-g}}\,
\bigg(
\ph \,\frac{\de S}{\de \ph}
- 2\, g_{\mu\nu}\,
\frac{\de S}{\de g_{\mu\nu}}
-\,B_{\mu\nu}\,\frac{\de S}{\de B_{\mu\nu}}
\bigg)\Ga_{\text{ind}}
\,=\,\langle \mathcal{T} \rangle\,.
\label{Gamaind}
\eeq
From the technical side, the simplest way to solve this equation
\cite{rie,frts84} is by changing the variables according to
(\ref{confBg}). This change reduces the sum of the three derivatives
in (\ref{Gamaind}) to a single variational derivative $\de/\de \si$.
After that, the nonlocal part of the covariant solution (there is also
local non-covariant version in terms of $\si$, which is obtained
even easier) can be obtained by using the conformal identity
for the modified topological term
\beq
&&
\,\,
\sqrt{-g}\,\biggl(E_4-\frac23\,{\cx} R\biggr)
\,=\,
\sqrt{-\bar{g}}\,\biggl({\bar E_4}-\frac23\,{\bar \cx} {\bar R}
+ 4{\bar \De_4}\si \biggr),
\label{119}
\\
&&
\mbox{where}
\quad\,\,
\Delta_4 \,=\,
\cx^2 + 2R^{\mu\nu}\nabla_{\mu}\nabla_{\nu} - \dfrac{2}{3}R\cx
+\dfrac{1}{3}(\nabla^{\mu}R)\nabla_{\mu}
\eeq
is the conformal Paneitz operator \cite{FrTs-superconf,Paneitz},
$\sqrt{-g}\Delta_4=\sqrt{-\bar{g}}\bar{\Delta}_4$.

The solution for the nonlocal part has the general form, which
does not depend on the form of the conformal terms $Y$
(see, e.g., \cite{AtA,Cheshire2} or \cite{OUP} for full detail),
\beq
&&
\Ga_{\text{ind},\,\text{nonloc}}\,\,=\,\,
\frac{b}{8}\int_x\int_y \biggl( E_4
-\frac23\square R \biggr)_{\hspace{-1mm}x}
G(x,y)\biggl(E_4-\frac23\square R\biggr)_{\hspace{-1mm}y}
\nn
\\
&&
\qquad \qquad \qquad
+\,\,
\frac{1}{4}\int_x\int_y Y(x)\, G(x,y)
\biggl(E_4-\frac23\square R\biggr)_{\hspace{-1mm}y}\,,
\label{nonlocal-Ga}
\eeq
where we used $\int_x\equiv \int \rd^4 x\sqrt{-g(x)}$ and the Green
function $G$ is of the Paneitz operator
\beq
(\sqrt{-g}\De_4)_x\,G(x,y)\,=\,\de(x,y).
\label{Green_function}
\eeq
Finally, the local part of the induced effective action results from
the integration
of the total derivative terms. These terms are, typically, subject to
ambiguities. In the present case, those are owing to the choice
of regularization (dimensional or higher-derivative Pauli-Villars
versions may produce different results \cite{anomaly-2004}, or
because of the different schemes of doubling in the fermionic
sector \cite{Cheshire2}. Anyway, since there are no mixed
scalar-$B_{\mu\nu}$ total derivative terms in the anomaly, we can
use the known results from \cite{AnoIntScal} and \cite{Cheshire2}.
In the $B_{\mu\nu}$-sector we get
\beq
&&
\Ga^{(1)}_{\text{ind},\,\ga_1}
\,\,=\,\,
-\,\, \frac{4s\,\ga_1}{3(4\pi)^2} \int_x
\,R B_{\mu\nu}B^{\mu\nu}\,,
\label{Gaga1}
\\
&&
\Ga^{(1)}_{\text{ind},\,\ga_2}
\,\,=\,\,
\frac{4s\,\ga_2}{3(4\pi)^2} \int_x 
\,\Big\{3\, \big( \na_\al B_{\mu\nu}\big)^2
\,-\,2\,R B_{\mu\nu}B^{\mu\nu}\Big\}\,,
\label{ga1ga2}
\eeq
Here
\beq
&&
\ga_1\,=\,0\,,
\,\,
\ga_2\,=\,1
\quad
\mbox{or }
\quad
\ga_1\,=\,1\,,
\,\, \ga_2\,=\,0
\label{ga2}
\eeq
for the schemes of doubling used in Sec.~\ref{sec3} or
in Appendix B, respectively.

In the scalar and purely gravitational sectors, we get
\beq
&&
\Ga_{\text{ind},\,\text{loc}}\,\,=\,\,
- \,\frac{\be_\tau }6\,\int_x R\ph^2
\,-\, \frac{3c-2b}{36}\,\int_x R^2 \,.
\label{local-Ga-ph}
\eeq
The overall expression for the anomaly-induced effective action
is the sum of (\ref{nonlocal-Ga}), (\ref{local-Ga-ph}), and
(\ref{ga1ga2}). This action preserves all valuable information
about the UV limit of the one-loop corrections and presents it
in a compact and useful form.

\subsection{Low-energy limit and effective potential}
\label{sec53}

The remaining part is to consider the IR limit of the anomaly-induced
action. Following \cite{AnoIntScal,AtA} we take this limit in a way
that is a standard one in general relativity, supplemented by some
conditions for the background scalar and antisymmetric field.

In the first place, the IR limit implies that the gravitational field
is weak. Using the parametrization
$g_{\mu\nu} = \eta_{\mu\nu} + h_{\mu\nu}$, we assume that
$\big|h_{\mu\nu}\big| \ll 1$, such that, in particular,
\beq
|R^2_{\mu\nu\al\be}| \ll |\cx R|,
\qquad
|R^2_{\mu\nu}| \ll |\cx R|,
\quad
\mbox{and}
\quad |R^2\ll |\cx R|.
\eeq

Secondly, we assume that scalar and $B_{\mu\nu}$-dependent
terms dominate over the terms with metric derivatives,
\beq
\big| \ph^2\big|  \gg \big|R_{\ldotp\ldotp\ldotp\ldotp}\big|\,,
\qquad
\big|(\nabla\ph)^2\big| \gg \big|R^2_{\ldotp\ldotp\ldotp\ldotp}\big|\,,
\qquad
\big|B^2_{\mu\nu}\big| \gg \big|R_{\ldotp\ldotp\ldotp\ldotp}\big|\,.
\label{IR-1}
\eeq

These conditions produce an essential reduction in the
anomaly-induced action. The simplifications in the local terms
are obvious, so let us concentrate on the nonlocal part given by
(\ref{nonlocal-Ga}). The Green function of the Paneitz operator
boils down to
\beq
G\,=\,\Delta^{-1}_4
\,\,\longrightarrow\,\,
\frac{1\,}{\,\cx^2}
\label{apprG}
\eeq
and the ``corrected'' topological term to
\beq
&&
E_4 - \frac{2}{3} \cx R
\,\,\longrightarrow\,\,
-\,\frac23\,\cx R\,.
\label{E4red}
\eeq
As a result, the first term in (\ref{nonlocal-Ga}) becomes an
addition $(b/18)\int_xR^2$ to the irrelevant (in our approximation
scheme) local term in (\ref{local-Ga-ph}). Furthermore, we meet
the reduction
\beq
&&
E_4 - \frac{2}{3} \cx R + \frac{1}{b}Y
\,\,\longrightarrow \,\,
-\,\frac23\,\cx R
\,+\, \frac{1}{b}\,Y_{\text{red}}\,,
\quad
\mbox{where}
\quad
Y_{\text{red}} \,=\, Y\Big|_{\om \to 0}\,.
\label{Yred}
\eeq
After a small algebra, the IR remnant of the nonlocal part of
induced action (\ref{nonlocal-Ga}) becomes
\beq
&&
\Gamma_{\text{ind}, \, \text{nonloc}} \,\, \approx \,\,-\,\,
\frac{1}{6}\int_x\int_y
\big(Y_{\text{red}}\big)_x\,\,
\biggl( \frac{1}{\cx} \biggr)_{x,y}\,\, \big(R \big)_y
\,\,+\,\, \mbox{local terms}.
\label{Ga-ind-IR}
\eeq
The scalar part of this expression was explored in \cite{AnoIntScal}
for the Abelian model, but the difference with the $SU(2)$ case is
small, hence it makes no sense to repeat the respective discussion.
Let us only repeat the main result. Using conformal parametrization
(\ref{confBg}) and the identification
\beq
&&
\si \,\longrightarrow \,-\,\ln \big( \ph/\bar{\ph}\big),
\label{param}
\eeq
the scalar part of the expression (\ref{Ga-ind-IR}) boils down to
the conventional one-loop effective potential,
\beq
V_{\text{eff}}^{(1)}(\ph)
\,\,=\,\,
\frac{1}{4!}\,\biggl(
\lambda\,+\,\frac12\,\widetilde{\beta}_f
\ln  \frac{\ph^2}{\mu^2} \biggr) \ph^4
\,\,-\,\,
\frac{1}{12}
\biggl( 1\,+\,  \ga \ln
\frac{\ph^2}{\mu^2}\biggr)R\ph^2,
\label{VEfPo}
\eeq
where $\widetilde{\be}_f =\be_f + 4f\ga$ and we identified
$\bar{\ph}$ with the conventional renormalization parameter $\mu$.
The standard form of the expression (\ref{VEfPo}) confirms that
the anomaly-induced action is a version of a renormalization group
improved classical action in the Minimal Subtraction scheme of
renormalization in curved spacetime. The main difference with the
renormalization group approach is that the conformal factor $\si$
depends on the spacetime coordinates, while the corresponding
renormalization group parameter is a constant.

In the purely $B_{\mu\nu}$-dependent sector, we can change the
identification from (\ref{param}) to
\beq
&&
\sigma \,\longrightarrow \,-\,\frac12\,
\ln \big(B^2_{\mu\nu}/ \bar{B}_{\mu\nu}^2\big).
\label{paramB}
\eeq
Both relations (\ref{param}) and (\ref{paramB}) are using the
transformations (\ref{confBg}) with the fiducial quantities
$\bar{\ph}$ and $\bar{B}_{\mu\nu}$ playing the role of
the renormalization parameter $\mu$.

Taking Eq.~(\ref{Ga-ind-IR}) in the linear in $\sigma$
approximation, we get at the leading terms in the form
\beq
&&
\frac{1}{\cx}\,=\,e^{2\sigma}\frac{1}{\bar{\cx}},
\qquad
R\,=\,e^{-2\si}\big[\bar{R}-6\bar{\cx}\sigma+O(\sigma^2)\big],
\label{R-exp}
\eeq
where
$\bar{\cx}=\bar{g}^{\mu\nu}\partial_\mu\partial_\nu$. Assuming
weak fiducial gravitational field $\bar{g}_{\mu\nu}$, we regard
$\bar{R}$ negligible. Then the factors $\,1/\bar{\cx}\,$ and
$\,\bar{\cx}\,$ cancel out and the product of the last two factors
becomes a delta function. After integration, we arrive at the
one-loop corrected potential of $B_{\mu\nu}$ in (\ref{actB}),
\beq
&&
V_{\text{eff}}^{(1)}(B)
\,\,=\,\,
- \frac12 \bigg[ \tau
+ \frac12 \,\be_\tau \ln \Big(B^2_{\mu\nu}/ \bar{B}_{\mu\nu}^2\Big)
\bigg] W_4
- \frac12 \bigg[ \la
+ \frac12 \,\be_\la \ln \Big(B^2_{\mu\nu}/ \bar{B}_{\mu\nu}^2\Big)
\bigg]W_1
\nn
\\
&&
\qquad \qquad \quad
+ \,\, \frac14 \bigg[f_2 + \frac12 \,\be_{f_2}
\ln \Big(B^2_{\mu\nu}/ \bar{B}_{\mu\nu}^2\Big) \bigg] W_2
+ \frac14 \bigg[f_3 + \frac12 \,\be_{f_3}
\ln \Big(B^2_{\mu\nu}/ \bar{B}_{\mu\nu}^2\Big) \bigg] W_3,
 \qquad \quad
\label{VB}
\eeq
It is worth noting that the term with $W_1$ in Eq.~(\ref{VB})
is presumably small in the described approximation. So, we
included it here only because this can be done without real effort.
Another detail is that the effective potential (\ref{VB}) differs
from the one obtained recently in \cite{Petrov}, because the last
comes from the non-renormalizable interaction of a quantum
fermion with an antisymmetric tensor field.

In the most complete case, when both scalar and $B_{\mu\nu}$
fields are present, the identification of the variable scale can be
done according to (\ref{param}), or (\ref{paramB}), or using, e.g.,
a linear combination of
$\ph^2$ and $B_{\mu\nu}^2$.\footnote{It is important to note that
the form of the operators $\hat{P}$ and $\hat{S}_{\al\be}$, as
quoted in the Appendices A and B, do not hint towards the most
physical or natural identification. In case of a direct derivation of
the potentials, there will be distinct logarithms in the different
sectors of the potential.}
The changes in the ``pure'' potentials (\ref{VEfPo}) and (\ref{VB})
reduce to the simple replacement of the logarithmic terms.

Assuming, for the sake of definiteness, the scale identification
(\ref{param}), the remaining ``mixed'' part of effective potential
has the form
\beq
&&
V_{\text{eff}}^{(1)}(\ph,B)
\,\,=\,\,
- \frac12 \bigg[ \xi_2
+ \frac12 \,\be_{\xi_2 } \ln \Big(\frac{\ph}{\bar{\ph}}\Big)\bigg]
B_{\mu \nu}^2\, \ph^2\,.
\label{Vxi2}
\eeq

The expressions for the effective potentials  (\ref{VEfPo}),
(\ref{VB}) and (\ref{Vxi2}) may be obtained by solving the
renormalization group equations based on the Minimal
Subtraction Scheme of renormalization and the scale identification
\cite{nelspan82,tmf,book}. In this case, there will be the same
argument of logarithms in all three cases, something that we can
easily provide by using the approach based on anomaly. This analogy
shows that the anomaly-induced action in general, and its IR
part, that can be linked to the effective potential, is nothing but
the local version of the renormalization group, when the global
parameter of metric rescaling is replaced by the local parameter,
i.e., the function $\si(x)$.

One can note that the choice of the arguments of logarithms in all
three expressions (\ref{VEfPo}), (\ref{VB}) and (\ref{Vxi2}) is
ambiguous, as always in the expressions restored from the Minimal
Subtraction Scheme. It is important to note that
the form of the operators $\hat{P}$ and $\hat{S}_{\al\be}$, as
quoted in the Appendices A and B, do not hint towards the most
physical or natural identification. In the case of a direct derivation
of the potentials, there will be distinct logarithms in the different
sectors of the potential.

\section{Conclusions}
\label{Conc}

We explored the renormalization in quantum theory of interacting
fields on the background of the metric and antisymmetric tensor field
$B_{\mu\nu}$. The model under consideration was quite general,
with the presence of Dirac fermions, scalars and gauge vectors. The
symmetry group considered here was  $SU(2)$, but most of the results
are universal and not expected to modify under the change of
symmetry group or representation of the fields.

In the previous works on the subject \cite{Avdeev1993,Cheshire}, it
was shown that the renormalizable interaction of the classical
$B_{\mu\nu}$ with quantum fermions requires vacuum action of
$B_{\mu\nu}$ which is different from the gauge-invariant Kalb-Ramon
model \cite{OgiPolu67,KalbRamon}. Instead, this vacuum action has to
follow local conformal symmetry, even in the case of massive fermions,
which are not conformal. The reason is that the mass term does not
violate the conformal symmetry in the kinetic terms. Here we extend
the formulation of a renormalizable theory on the $B_{\mu\nu}$
background to the interacting fields. In particular, we show that
renormalizability requires not only fermions, but also scalars to
have nonminimal interaction to $B_{\mu\nu}$, similar to the case of
quantum field theory with torsion \cite{bush85,torsi}.

The renormalization group equations for the nonminimal effective
charges corresponding to the interaction of fermions and scalars with
$B_{\mu\nu}$ show that the corresponding interactions become
stronger in the UV limit. One can show that this result does not
depend on the gauge group and, therefore, is expected to hold in
any interacting theory with the Yukawa interaction.

Finally, we derive the trace and anomaly-induced effective action
$B_{\mu\nu}$, metric and scalar field. Taking the IR limit
in a way proposed recently in
\cite{AnoIntScal,AtA} and \cite{Vale2023}, we arrive at the
effective potential for scalar and $B_{\mu\nu}$ fields. In principle,
such a potential may be further explored, including in relation to
possible physical applications, as discussed, e.g., in \cite{Petrov}.

\section*{Acknowledgements}

T.M.S. is grateful to Funda\c{c}\~{a}o de Amparo \`{a} Pesquisa
do Estado de Minas Gerais (FAPEMIG) for supporting his MSc project.
I.Sh. is grateful to CNPq (Conselho Nacional de Desenvolvimento
Cient\'{i}fico e Tecnol\'{o}gico, Brazil)  for the partial support
under the grant 305122/2023-1.

\section*{Appendix A}
\label{ApA}

The intermediate formulas for the elements of the operator
(\ref{opmin}) and the derivation of divergences (\ref{GaBvac}),
include
\beq
\mathbf{1} =
\begin{pmatrix}
            1 & 0 & 0 \\
            0 & \de^{ab} \de^{\nu}_{\mu} & 0\\
            0 & 0 & \de^{ab} 1
\end{pmatrix},
\qquad
\mathbf{\Pi} =
        \begin{pmatrix}
            \Pi_{11} & \Pi_{12} & \Pi_{13} \\
            \Pi_{21} & \Pi_{22} & \Pi_{23} \\
            \Pi_{31} & \Pi_{32} & \Pi_{33}
        \end{pmatrix},
\eeq
where
\beq
\Pi_{11} &=&
( m_s^2 - \xi_1 R - \xi_2 B_{\mu \nu}^{2} ) \de^{ab}
+ \frac{f}{6} ( \ph^2 \de^{ab} + 2 \ph^{a} \ph^{b} ),
\nn
\\
\Pi_{12} &=&
2 g \vp^{acb} ( \na^{\nu} \ph^{c} ),
\qquad
\Pi_{13}
\,=\, - i m_{f} h \vp^{acb} \bar{\psi}_{l}^{c},
\nn
\\
\Pi_{21}
&=& g \vp^{acb} ( \na_{\mu} \ph^{c} ),
\qquad \,\,\,
\Pi_{22} \,=\,
-R^{\nu}_{\mu} \de^{ab} + g^{2} ( \ph^2 \de^{ab}
- \ph^{a} \ph^{b} ) \de^{\nu}_{\mu},
\nn
\\
\Pi_{23}
&=&
i m_{f} g \vp^{acb} \bar{\psi}_{l}^{c} \ga_{\mu},
\quad \,\,\,\,\,
\Pi_{31} = - 2 i h \vp^{acb} \psi_{k}^{c},
\qquad \Pi_{32} = - 2i g \vp^{acb} \ga^{\nu} \psi_{k}^{c},
\nn
\\
\Pi_{33}
&=&
\de_{kl} \Big[
\Big( m_{f}^{2}
- \frac{1}{4} R + i \eta m_f  B_{\mu \nu} \Si^{\mu \nu} \Big)\de^{ab}
+ i h m_f \vp^{acb} \ph^{c} \Big],
\eeq
and
\beq
\mathbf{h}^\al
\,\,=\,\,
\frac12
\begin{pmatrix}
0
&  g \vp^{acb} \ph^{c} g^{\nu \al}
& h \vp^{acb} \bar{\psi}_{l}^{c} \ga^{\al}
\\
- g \vp^{acb} \ph^{c} \de^{\al}_{\mu}
&
0
& - g \vp^{acb} \bar{\psi}_{l}^{c} \ga_{\mu} \ga^{\al}
\\
0
&
0
& - \de_{kl} \big( h \vp^{acb} \ph^{c}
+ \eta B_{\mu \nu}
\Si^{\mu \nu} \de^{ab}  \big) \ga^\al
\end{pmatrix},
\eeq
The last term in this matrix can be reduced, but we use this
form for brevity.

Furthermore,
\beq
\mathbf{P} =
\begin{pmatrix}
            P_{11} & P_{12} & P_{13} \\
            P_{21} & P_{22} & P_{23} \\
            P_{31} & P_{32} & P_{33}
\end{pmatrix},
\qquad
\mathbf{S}_{\al \be} \,=\,
        \begin{pmatrix}
            0 & S_{12} & S_{13}  \\
            S_{21} & S_{22} & S_{23} \\
            0 & 0 & S_{33}
\end{pmatrix},
\eeq
where
\beq
P_{11}
&=&
\big(m_s^2 - \tilde{\xi}_1 R - \xi_2 B_{\mu \nu}^2 \big) \de^{ab}
+ \Big(\frac{f}{6} - g^2\Big)
\big(\ph^2\de^{ab} - \ph^{a} \ph^{b} \big),
    \nn
    \\
    P_{12} &=& \frac{3}{2} g \vp^{acb} ( \na^{\nu} \ph^{c} ),
    \nn
    \\
    P_{13} &=& -im_{f} h \vp^{acb} \bar{\psi}_{l}^{c} 
- \frac{1}{2} h \vp^{acb} ( \na_{\al} \bar{\psi}_{l}^{c} ) \ga^{\al} 
- g^{2} ( \de^{ab} \ph^{c} \bar{\psi}_{l}^{c} 
- \ph^{b}\bar{\psi}_l^a ) - h^2 (\de^{ab} \ph^c \bar{\psi}_l^c 
- \ph^{a} \bar{\psi}_{l}^{b} ),
    \nn
    \\
    P_{21} &=& \frac{3g}{2}\, \vp^{acb} ( \na_{\mu} \ph^{c} ),
\,\,    \qquad
P_{22} = \frac{1}{6} R \de^{ab} \de^{\nu}_{\mu}
- \de^{ab} R^{\nu}_{\mu}
+ \frac{3}{4} g^{2} ( \ph^2 \de^{ab}
- \ph^{a} \ph^{b} ) \de^{\nu}_{\mu},
\nn
\\
P_{23}
&=&
i g m_f \vp^{acb} \bar{\psi}_{l}^{c} \ga_{\mu}
+ \frac{1}{2} g \vp^{acb} ( \na_{\al} \bar{\psi}_{l}^{c} ) \ga_{\mu} \ga^{\al}
- \frac{1}{4} g h ( \de^{ab} \ph^{c} \bar{\psi}_{l}^{c} - \ph^{b} \bar{\psi}_{l}^{a} ) \ga_{\mu}
\nn
\\
&&
+ g h ( \de^{ab} \bar{\psi}_{l}^{c} \ph^{c} - \bar{\psi}_{l}^{b} \ph^{a} ) \ga_{\mu},
\nn
\\
P_{31} &=&
-2ih \vp^{acb} \psi_{k}^{c},
\quad \,\,\,\qquad
P_{32} = - 2 i g \vp^{acb} \ga^{\nu} \psi_{k}^{c},
\nn
\\
P_{33}
&=&
\de_{kl} \Bigl[ \de^{ab} \Big( m_f^2
- \frac{1}{12} R + i \eta  m_f B_{\mu \nu} \Si^{\mu \nu} \Big)
+ i h m_f \vp^{acb} \ph^{c}
+ \frac{1}{2} h \vp^{acb} ( \na_{\be} \ph^{c} ) \ga^{\al}
\nn
\\
&&
+ \eta ( \na_{\al} B_{\mu \nu} ) \de^{ab}
\Si^{\mu \nu} \ga^\al
+ h^2 ( \de^{ab} \ph^{2} - \ph^{a} \ph^{b} )
- \eta h B_{\mu \nu} \vp^{acb} \ph^{c} \Si^{\mu \nu} \Bigr]
\eeq
and
\beq
S_{12}
&=&
\frac{1}{2} g \vp^{acb} [ ( \na_{\be} \ph^{c} ) \de^{\nu}_{\al} 
- ( \na_{\al} \ph^{c} ) \de^{\nu}_{\be} ],
    \nn
    \\
S_{13}
&=&
\frac{1}{2} h \vp^{acb} [ \na_{\be} \bar{\psi}_{l}^{c} ) \ga_{\al}
- ( \na_{\al} \bar{\psi}_{l}^{c} ) \ga_{\be} ]
- \frac{1}{4} g^2 ( \ph^{c} \bar{\psi}_{l}^{c} \de^{ab}
- \bar{\psi}_{l}^{a} \ph^{b} ) ( \ga_{\be} \ga_{\al} - \ga_{\al} \ga_{\be} )
    \nn
    \\
&&
-\frac{1}{4} h^{2} ( \ph^{c} \ph^{c} \de^{ab}
- \ph^{a} \bar{\psi}_{l}^{b} ) ( \ga_{\be} \ga_{\al} - \ga_{\al} \ga_{\be} )
+ \frac{1}{4} \eta h
B_{\mu \nu} \vp^{acb} \bar{\psi}_{l}^{c}
\ga_{\be} \Si^{\mu \nu} \ga_{\al},
\nn
    \\
    S_{21} &=& -\frac{1}{2} g \vp^{acb} \left[ ( \na_{\be} \ph^{c} ) g_{\mu \al} - ( \na_{\al} \ph^{c} ) g_{\mu \be} \right],
    \nn
    \\
    S_{22} &=& \de^{ab} R^{\nu}{}_{\mu \al \be} + \frac{1}{4} g^{2} ( \ph^{c} \ph^{c} \de^{ab} - \ph^{a} \ph^{b} ) ( g_{\mu \be} \de^{\nu}_{\al} - g_{\mu \al} \de^{\nu}_{\be} ),
    \nn
    \\
    S_{23} &=& -\frac{1}{2} g \vp^{acb} \left[ ( \na_{\be} \bar{\psi}_{l}^{c} ) \ga_{\mu} \ga_{\al} - ( \na_{\al} \bar{\psi}_{l}^{c} ) \ga_{\mu} \ga_{\be} \right] - \frac{1}{4} g h ( \ph^{c} \bar{\psi}_{l}^{c} \de^{ab} - \ph^{b} \bar{\psi}_{l}^{a} ) ( g_{\mu \be} \ga_{\al} - g_{\mu \al} \ga_{\be} )
    \nn
    \\
    &&+\frac{1}{4} g h ( \bar{\psi}_{l}^{c} \ph^{c} - \ph^{a} \bar{\psi}_{l}^{b} ) \ga_{\mu} ( \ga_{\be} \ga_{\al} - \ga_{\al} \ga_{\be} ) - \frac{1}{4} g \eta B_{\mu \nu} \vp^{acb} \bar{\psi}_{l}^{c} \ga_{\mu} \ga_{\be} \Si^{\mu \nu} \ga_{\al},
    \nn
    \\
S_{33}
&=&
\de_{kl} \Bigl\{
 2 i \eta^{2} \de^{ab} \ga^{5} \big(
B_{\be \mu} \tilde{B}_{\al}{}^{\mu} - B_{\al \mu} \tilde{B}_{\be}{}^{\mu} \big)
+ \eta^2 \de^{ab} ( B_{\be \nu} B_{\al \mu}
+ \tilde{B}_{\be \nu} \tilde{B}_{\al \mu} )
( \ga^{\mu} \ga^{\nu} - \ga^{\nu} \ga^{\mu} )
 \nn
\\
&&
- \frac{\eta}{2} \,\de^{ab}
\big[ ( \na_{\be} B_{\mu \nu} \Si^{\mu \nu} \ga_{\al}
- ( \na_{\al} B_{\mu \nu} ) \Si^{\mu \nu} \ga_{\be} \big]
- \eta h \ph^{c} \vp^{acb} \ga^{5} \ga^{\mu}
(\tilde{B}_{\be \mu} \ga_{\al} - \tilde{B}_{\al \mu} \ga_{\be})
 \nn
\\
&&
- \frac{h}{2}\vp^{acb} \big[(\na_{\be} \ph^{c})\ga_{\al}
 - ( \na_{\al} \ph^{c} ) \ga_{\be} \big]
+ \frac{ih^2}{2}  ( \ph^2 \de^{ab} - \ph^{a} \ph^{b} )
\Si_{\al\be}
- \frac{1}{4} R_{\al \be \la \ta} \ga^{\la} \ga^{\ta} \de^{ab}
\Bigr\}.
\qquad
\eeq
In these expressions we used the notation
$\tilde{B}^{\al\be} = \frac12 \vp^{\al\be\mu\nu} B_{\mu\nu}$.

\section*{Appendix B}

In order to verify the calculations and ensure consistency, we consider
an alternative form for the doubling operator,
\beq
    \textbf{H}^* =
    \begin{pmatrix}
- \,\de^{ab}\,\,  &   0   &   0 \\
0 & \,\de^{ab} \de^{\nu}_{\mu} \, & 0 \\
0 & 0 & - \frac{i}2 \delta_{kl}
        \de^{ab}( \slashed{\na} - \eta B_{\mu \nu} \Si^{\mu \nu} - i m_f )
    \end{pmatrix},
\eeq
\beq
\mathbf{h}^{\al}
\,\,=\,\,
\begin{pmatrix}
0
& \frac{1}{2} g \vp^{acb} \ph^{c} g^{\nu \al}
& \frac{1}{2} h \vp^{acb} \bar{\psi}_{l}^{c} \ga^{\al}
\\
- \frac{1}{2} g \vp^{acb} \ph^{c} \de^{\al}_{\mu}
&
0
& - \frac{1}{2} g \vp^{acb} \bar{\psi}_{l}^{c} \ga_{\mu} \ga^{\al}
\\
0
&
0
& \de_{kl} \big( 2 \eta \de^{ab} \tilde{B}^{\al \be} \ga^{5} \ga_{\be}
- \frac{1}{2} h \ep^{acb} \ph^{c} \ga^{\al}
\end{pmatrix},
\eeq
The structures of $\mathbf{\hat{\Pi}}$, $\mathbf{\hat{P}}$ and
$\mathbf{\hat{S}}$ remain unchanged except for the following entries:
\beq
\Pi_{13} &=& - \eta h \ep^{acb} B_{\mu \nu} \bar{\psi}_{l}^{c}
\Si^{\mu \nu} - i m_{f} h \ep^{acb} \bar{\psi}_{l}^{c},
    \nn
    \\
\Pi_{23} &=&
2 i \eta g \ep^{acb} B_{\mu \nu} \bar{\psi}_{l}^{c} \ga^\nu
- 2 \eta g \ep^{acb} \tilde{B}_{\mu \nu} \bar{\psi}_{l}^{c} \ga^{5} \ga^{\nu},
\nn
\\
\Pi_{33}
&=&
\de_{kl} \biggl\{ \de^{ab}
\Bigl[ m_{f}^{2} - \frac{1}{4} R
- 2 i \eta ( \na_{\mu} B^{\mu \nu} ) \ga_{\nu}
+ 2 \eta ( \na_{\mu} \tilde{B}^{\mu \nu} ) \ga^{5} \ga_{\nu}
\nn
\\
&&
- 2 i \eta^{2} B_{\mu \nu} \tilde{B}^{\mu \nu} \ga^{5}
+ 2 \eta^2 B_{\mu \nu}^{2} \Bigr]
+ \eta h \ep^{acb} \ph^{c} B_{\mu \nu} \Si^{\mu \nu}
+ i m_{f} h \ep^{acb} \ph^{c} \biggr\},
\eeq
and also
\beq
&&
P_{33}
= \de_{kl} \Bigl[ \de^{ab} \Bigl( m_{f}^{2}
- \frac{1}{12} R - 2 i \eta ( \na_{\mu} B^{\mu \nu} ) \ga_{\nu}
- 2 i \eta^2 B_{\mu \nu} \tilde{B}^{\mu \nu} \ga^{5}
- 2 \eta^2 B_{\mu \nu}^{2} \Bigr)
\nn
\\
&&
\qquad
- \,\,\eta h \ep^{acb} \ph^{c} B_{\mu \nu} \Si^{\mu \nu}
+ i m_{f} h \ep^{acb} \ph^{c}
+ \frac{1}{2} h \ep^{acb} ( \na_{\al} \ph^{c} ) \ga^{\al}
+ h^2 ( \de^{ab} \ph^{2} - \ph^{a} \ph^{b} ) \Bigr],
\quad
\quad
\eeq
\beq
S_{13}
&=&
\frac{1}{4} g^2 ( \de^{ab} \ph_{c} \bar{\psi}_{l}^{c}
- \ph^{b} \bar{\psi}_{l}^{a} ) ( \ga_{\be} \ga_{\al}
- \ga_{\al} \ga_{\be} )
- 2 \eta h \ep^{acb} \tilde{B}_{\al \be} \bar{\psi}_{l}^{c} \ga^{5}
\nn
\\
&&
+ \frac{i \eta h}{2} \ep^{acb} ( \tilde{B}_{\al \nu}
\vp_{\be}{}^{\nu}{}_{\la \ta} - \tilde{B}_{\be \nu}
\varepsilon_{\al}{}^{\nu}{}_{\la \ta} )
\bar{\psi}_{l}^{c} \ga^{\la} \ga^{\ta}
+ \frac{1}{4} h^2 ( \de^{ab} \ph^{c} \bar{\psi}_{l}^{c}
- \ph^{a} \bar{\psi}_{l}^{b} )
    \nn
    \\
&&
+ \frac{1}{2} \ep^{acb} [ ( \na_{\be} \bar{\psi}_{l}^{c} ) \ga_{\al}
-  (\na_{\al} \bar{\psi}_{l}^{c} ) \ga_{\be} ],
\nn
\\
S_{23}
&=&
- \frac{1}{2} g \vp^{acb} [ ( \na_{\be} \bar{\psi}_{l}^{c} )
\ga_{\mu} \ga_{\al}
- ( \na_{\al} \bar{\psi}_{l}^{c} ) \ga_{\mu} \ga_{\be} ]
+ \frac{1}{4} g h ( \de^{ab} \ph^{c} \bar{\psi}_{l}^{c}
- \ph^{b} \bar{\psi}_{l}^{a} ) ( g_{\mu \be} \ga_{\al}
- g_{\mu \al} \ga_{\be} )
\nn
\\
&&
- \frac{1}{4} g h ( \ph^{c} \bar{\psi}_{l}^{c} \de^{ab}
- \ph^{a} \bar{\psi}_{l}^{b} ) \ga_{\mu} ( \ga_{\be} \ga_{\al}
- \ga_{\al} \ga_{\be})
- g \eta \vp^{acb}(\tilde{B}_{\al \mu} \bar{\psi}_{l}^{c}\ga_\be
- \tilde{B}_{\be \mu} \bar{\psi}_{l}^{c} \ga_{\al} ) \ga^{5}
\nn
\\
&&
+ 2g \eta \varepsilon^{acb} ( \tilde{B}_{\al \nu} g_{\mu \be}
- \tilde{B}_{\be \nu} g_{\mu \al} ) \ga^{\nu} \ga^{5}
+ 2g\eta\vp^{acb}\tilde{B}_{\al \be}\bar{\psi}_{l}^{c}\ga_\mu\ga^5
    \nn
    \\
&&
- i g \eta \vp^{acb} \big(
  \tilde{B}_{\al \,\,\cdot}^{\,\,\,\,\, \nu} \vp_{\be \nu\mu \la}
- \tilde{B}_{\be \,\,\cdot}^{\,\,\,\,\, \nu} \vp_{\al \nu\mu \la}
\big)
\bar{\psi}_{l}^{c} \ga^{\la}.
\eeq
The result for the divergences is the same as in the the first scheme
of doubling (except the total derivative terms which were discussed
in \cite{Cheshire2}), which is a strong confirmation of the correctness
of the calculations.

\section*{Appendix C}
\label{ApC}

The relations between bare and renormalized quantities are as
follows. For the fields,
\beq
\psi_0^{a}
    &=&
    \mu^{\frac{n-4}{2}} \biggl( 1 + \frac{h^2 + 2 g^2}{\ep} \biggr) \psi^{a},
    \nn
    \\
\ph_0^{a}
&=&
\mu^{\frac{n-4}{2}} \biggl(1 + \frac{4sh^2 - 4g^2}{\ep}\biggr)\ph^a.
\label{renBpsiph}
\eeq
For the nonminimal parameter $\eta$, we get
\beq
    \eta_{0} = \left( 1 - \frac{2h^2 + 4 g^2}{\ep} \right) \eta.
\eeq
For the couplings $h$, $f$ and nonminimal parameter $\xi_{2}$,
we have
\beq
&&
h_0 = \mu^{\frac{4-n}{2}}
\Big[ h - \frac{1}{\ep}\big(4 h^3 + 12 h g^2 + 4 s h^3 \big)\Big],
\nn
\\
\nn
\\
&&
f_0 = \mu^{4-n}
\Big[ f - \frac{1}{\ep}
\Big( \frac{11}{3} f^2 - 24 g^2 f + 72 g^4 + 16 s f h^2 - 96 s h^4
\Big) \Big]
\eeq
and
\beq
\xi_{2}^{0}
\,=\,
\xi_{2} + \frac{1}{\ep}
\biggl[ \biggl( 12 g^2 - 8 s h^2 - \frac{5}{3} f \biggr) \xi_{2}
+ 32 s \eta^2 h^2 \biggr].
\eeq


\end{document}